\documentclass[showpacs,preprintnumbers,amsmath,amssymb,11pt]{revtex4}
\usepackage{graphicx}
\usepackage{dcolumn}
\usepackage{bm}

\pdfoutput=1
\begin{document}

\title{Local and Nonlocal Dispersive Turbulence}
\author
{Jai Sukhatme$^1$ and Leslie M. Smith$^{1,2}$\\
   Mathematics Department, University of Wisconsin-Madison, Madison, WI 53706 \\
   Engineering Physics Department, University of Wisconsin-Madison, Madison, WI 53706 \\}
\date{\today}
\begin{abstract}

We consider the evolution of a family of two-dimensional (2D) dispersive turbulence models.
The members of this family involve the nonlinear advection of a dynamically active scalar field, and 
as per convention, the locality of the streamfunction-scalar relation is
denoted by $\alpha$, with smaller $\alpha$ implying increased locality ($\alpha = 1$ gives
traditional 2D dynamics).
The dispersive nature arises via a linear term whose strength, after 
non-dimensionalization, is characterized by a parameter $\epsilon$. Setting $0 < \epsilon \le 1$,
we investigate the
interplay of advection and dispersion for differing degrees of locality. 
Specifically, we study the forward (inverse) transfer of enstrophy (energy) under large-scale 
(small-scale)
random forcing along with the geometry of the scalar field. 
Straightforward arguments suggest that for small $\alpha$
the scalar field should consist of progressively larger isotropic eddies, while for large 
$\alpha$ the scalar field is expected to 
have a filamentary structure resulting from a stretch and fold mechanism; 
much like that of a small-scale passive field when advected by a large-scale smooth flow. 
Confirming this, we proceed to forced/dissipative dispersive numerical experiments 
under weakly non-local to local conditions (i.e.\ $\alpha \le 1$).
In all cases we see the establishment of well-defined
spectral scaling regimes. For $\epsilon \sim 1$, there is quantitative agreement between non-dispersive 
estimates and observed slopes in the inverse energy transfer regime. 
On the other hand, forward enstrophy transfer 
regime always yields slopes that are significantly 
steeper than the corresponding non-dispersive estimate. 
At present resolution, additional simulations show the scaling in the inverse regime to 
be sensitive to the strength of the dispersive term : specifically, as $\epsilon$ decreases, 
quite expectedly 
the inertial-range shortens but we also observe that the slope of the power-law decreases. On the other hand,
for the same range of $\epsilon$ values, 
the forward regime scaling is observed to be fairly universal.

\end{abstract}
\pacs{47.52.+j}

\maketitle

\section{Introduction}

It is quite common in geophysical fluid dynamics to encounter problems that involve the presence 
of both advection and dispersion (see for example, Chapter 5 in the 
text by Majda \cite{Majda-book} and Chapter 5 in the text by Chemin et\ al.\ \cite{Chemin-book}).
In a two-dimensional (2D) context, simple model equations that possess advection and dispersion
include 
the familiar barotropic beta plane equations \cite{Rhines} and the dispersive surface quasigeostrophic (SQG) 
equations \cite{HPGS}. An interesting aspect of such equations, highlighted by 
Rhines \cite{Rhines} in the barotropic beta plane case, is the co-existence of turbulent and wave-like motions 
\cite{HH}, \cite{Shep}. 
In fact,
the dispersive SQG equations have been proposed as an alternate (more local) 
platform for investigating wave-turbulence interactions \cite{HPGS}. \\

In the present work, we look at an extended family (that includes the aforementioned examples as members) of dispersive 
active scalars. This is a  
simple dispersive generalization of the family of 2D turbulence 
models introduced in Pierrehumbert, Held \& Swanson \cite{PHS}. Specifically, in a 
2D periodic setting we consider

\begin{eqnarray}
\frac{D \theta}{D t} + \frac{1}{\epsilon} \frac{\partial \psi}{\partial x} = 0 ~;~ (u,v)=(-\frac{\partial \psi}{\partial y},
\frac{\partial \psi}{\partial x}) \nonumber \\
~\textrm{where}~~ \hat{\theta}(k_x,k_y,t) = - (k^2)^\alpha \hat{\psi}(k_x,k_y,t)
\label{1}
\end{eqnarray}
where $D/Dt$ denotes the 2D material derivative and $\epsilon$ is a non-dimensional parameter. 
Note that in real space 
$\theta$ and $\psi$ above are related via a suitable (pseudo-) differential operator, i.e.\ $\theta = - (-\triangle)^{\alpha}
\psi$ where $\triangle$ is the 2D Laplacian, and for our purposes $\alpha \in \Re^{+}$.
For the 
beta plane equations we have $\alpha=1$, $\theta$ is the vorticity field, and $\epsilon$
corresponds to the Rhines number $Rh=U/(\beta L^2)$. In the dispersive SQG case $\alpha=1/2$, $\theta$ 
represents the buoyancy (or potential temperature \cite{HPGS}), 
and $\epsilon=U/(\Lambda L)$. Physically, of course we are dealing with very different scenarios 
wherein $\beta$ is the ambient planetary gradient of the vorticity
while $\Lambda$ is the background surface buoyancy gradient. Apart from $\alpha=1,1/2$, there exist 
other members of this family that have physical interpretations \cite{Smithetal}. But, from a broader
perspective, the entire family is well defined, and we expect that varying $\alpha$ would provide 
greater insight into the interplay between advection and dispersion. \\

We begin examining the effect of $\alpha$ on the dynamics by suppressing the dispersive term. After
developing a feel for the dependence of the energy/enstrophy transfer and the 
geometry of the scalar field on $\alpha$, we introduce 
dispersion and note its modulating effect on the nonlinear term. In the limit 
$\epsilon \rightarrow 0$, the so-called limiting dynamics are controlled by resonant interactions. In fact, the 
resonant skeleton corresponding to (\ref{1}) is seen to be of the same form as that of the 
rotating three-dimensional (3D) Euler equations. Following the original argument by Rhines \cite{Rhines},
for $0 < \epsilon \le 1$, 
if $\alpha > 0.5$, 
the simultaneous
transfer of energy to large scales and small frequencies leads to an anisotropic
streamfunction (resulting in predominantly
zonal flow). In contrast, given the nature of the dispersion relation, 
it turns out that for $\alpha < 0.5$ the
constraints of energy transfer to large scales and small frequencies do not require the 
spontaneous development
of anisotropy in
the streamfunction. Confirming this via decaying runs from 
spatially un-correlated initial data, we
proceed to a suite of forced/dissipative numerical simulations.
In all cases, under weakly-local to local conditions (i.e. $\alpha \le 1$) we 
note the formation of clear power-law scaling. The various slopes are extracted and compared with 
non-dispersive estimates. The paper ends with a collection of 
results and a brief discussion of potentially interesting avenues for future work. 

\section{Some observations on the effect of $\alpha$}

\begin{figure}
\centering
\includegraphics[width=4.5cm,height=4cm]{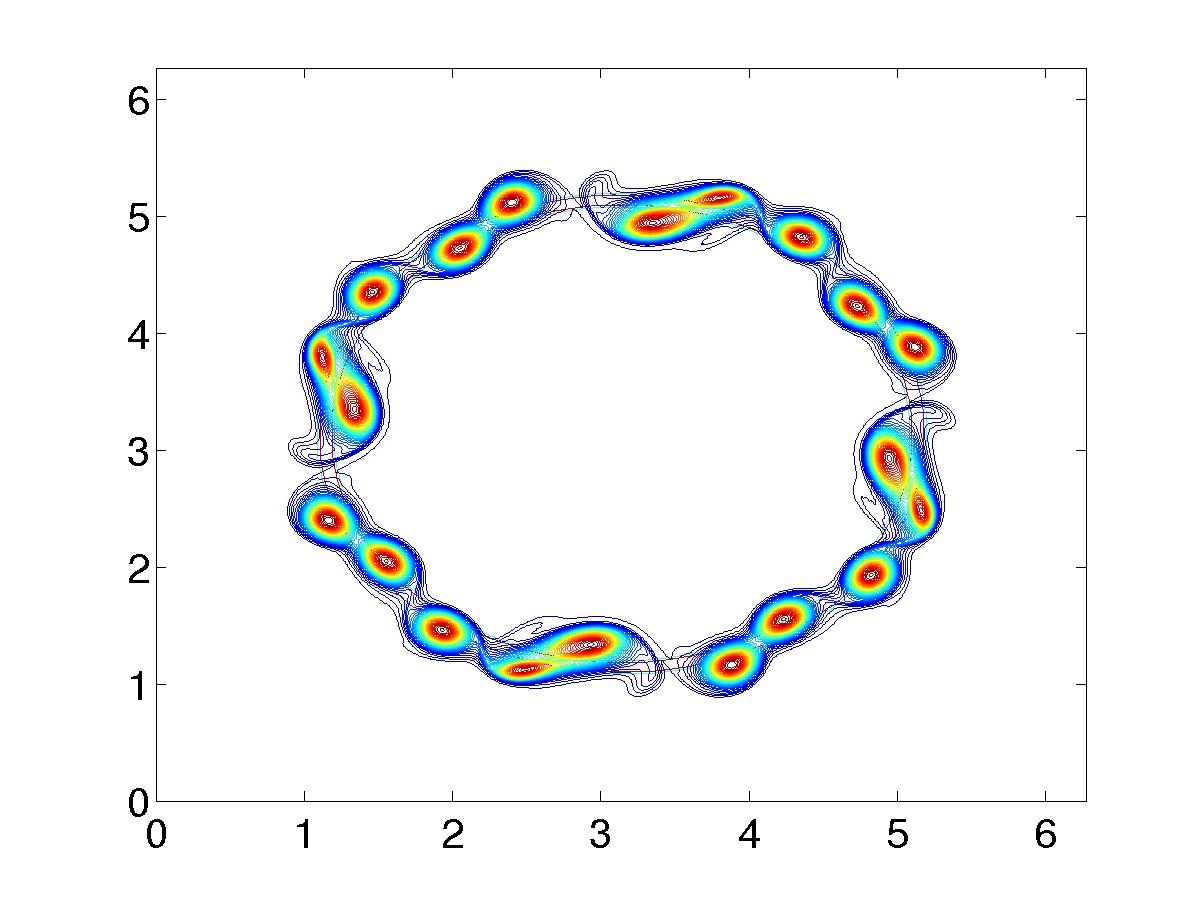}
\includegraphics[width=4.5cm,height=4cm]{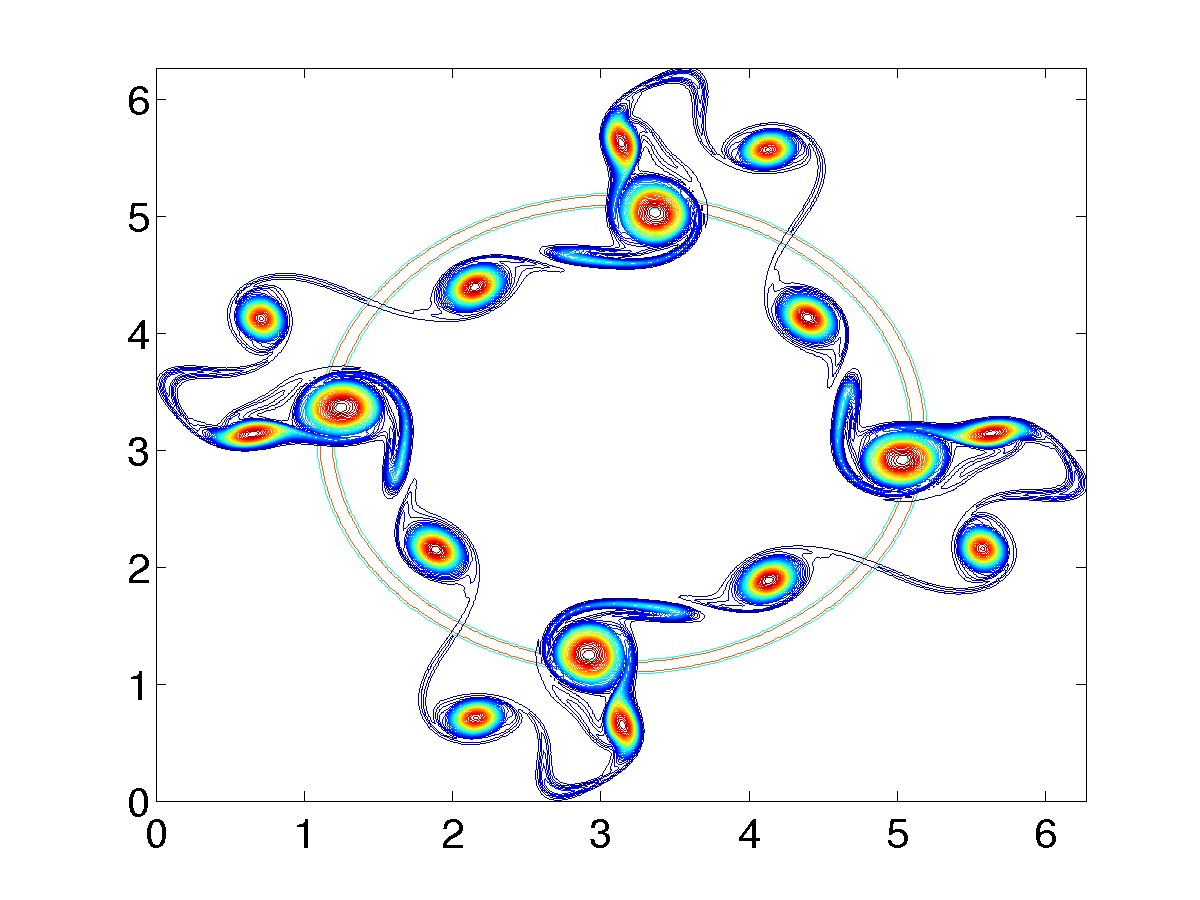}
\includegraphics[width=4.5cm,height=4cm]{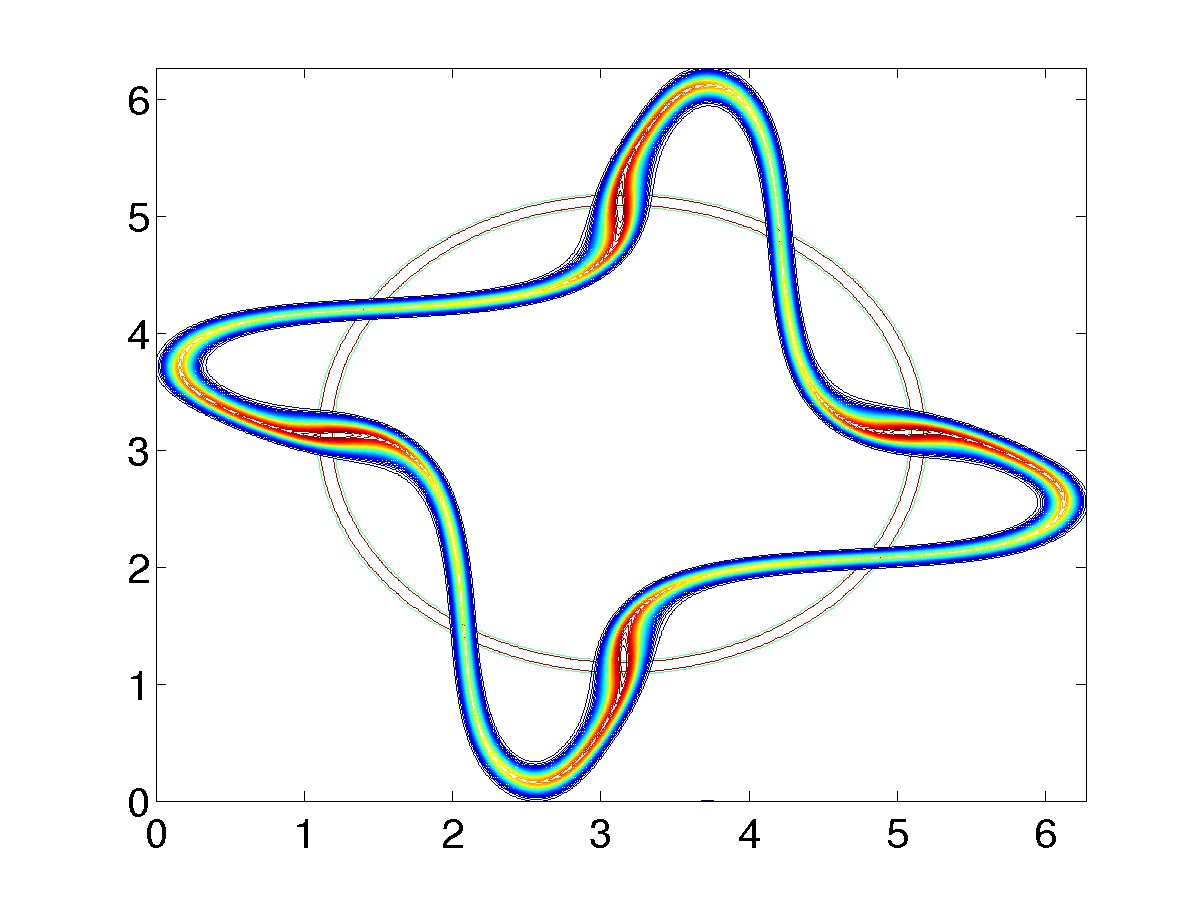}
\caption{\label{fig0} Non-dispersive evolution of $\theta$ rings. From 
left to right, $\alpha=0.5,1$ and 1.5 respectively. Quite clearly, for smaller $\alpha$ 
the deformation 
is more local in character. } 
\end{figure}

Before considering the effect of dispersion, we examine the influence of $\alpha$ in greater detail.
To get
a feel for the locality of the interactions, we shut off the linear dispersive term in (\ref{1}) and 
consider a simple numerical exercise involving the evolution of a smooth $\theta$ ring. 
As is seen in Fig. (\ref{fig0}), 
the deformation of the ring is more physically local for smaller $\alpha$.
Given that $\hat{\psi}(\vec{k})=-\hat{\theta}(\vec{k})/k^{2\alpha}$, 
as $\alpha$ increases, one sees that only the 
small $k$ features of 
$\theta$ and $\psi$ remain dynamically active. 
As a result, for larger
$\alpha$, smaller scale features of the scalar field are for all purposes driven in a passive manner.
In fact, in the limit $\alpha \rightarrow \infty$, 
only the wavenumber one ($k=1$)
component of $\theta$ and $\psi$ is coupled, and
we end up with the problem of passive advection via a large-scale smooth
flow. In contrast, the case of small $\alpha$ is markedly 
different. Specifically, $(\hat{u},\hat{v})=
({\rm i}k_y\hat{\theta}/k^{2\alpha},-{\rm i}k_x\hat{\theta}/k^{2\alpha})$ results in a transition 
at $\alpha =1/2$ (when the scalar and velocity fields have similar scales), where 
for $\alpha < 1/2$, the velocity fields
are in fact of a finer scale than the advected scalar field. Note that $\psi$ remains a smoothened 
form of $\theta$. We expect this transition to have interesting consequences 
as there are well-known differences in the 
behavior of a passive field when advected by large-scale, comparable-scale and small-scale flows 
respectively \cite{SP},\cite{FH},\cite{MajdaKramer}. \\

In a 2D periodic domain, much like its non-dispersive counterpart, (\ref{1}) conserves 
energy $E= \int_{\bf D} \psi \theta$ and enstrophy or "$\theta$-variance" 
${\mathcal E}=\int_{\bf D} \theta^2$. 
Continuing with the 
non-dispersive case, following arguments 
for the 2D Euler equations \cite{Kraichnan1967}, it is expected that the energy primarily flows 
to large scales while the $\theta$-variance is transferred to small scales. Assuming the 
presence of an equilibrium cascade, detailed 
spectral scaling laws in the appropriate inertial-range have also been derived \cite{PHS},
\cite{Smithetal}, \cite{Watanabe} (see also Tran \cite{Tran1} and Tran \& Bowman \cite{Tran2} for bounds on the 
scaling exponents for certain special dissipative operators in the SQG case). 
If we consider the Fjortoft estimate (\cite{Fjortoft}, Chapter 4 in Salmon 
\cite{Salmon-book}) i.e.\
the transfer of energy out of scale $k_1$ into scales $k_0,k_2$ 
(s.t.\ $k_0=k_1/2a$ and $k_2=2a k_1$, where $a > 1$), 
we have 
$E_0=E_1 \times \{ a^{2\alpha}/[1+a^{2\alpha}] \}$; i.e.
the more nonlocal 
the situation ($\alpha > 0.5$), the larger 
is the fraction of the energy (enstrophy) transferred to large (small) scales. 
Note that increasing $a$ shows this unequal distribution of energy and enstrophy is 
further exaggerated when the exchange
involves scales that are progressively further apart, i.e.\ 
the energy/enstrophy partition is more biased in spectrally nonlocal transfers. \\

\begin{figure}
\centering
\includegraphics[width=5cm,height=4.5cm]{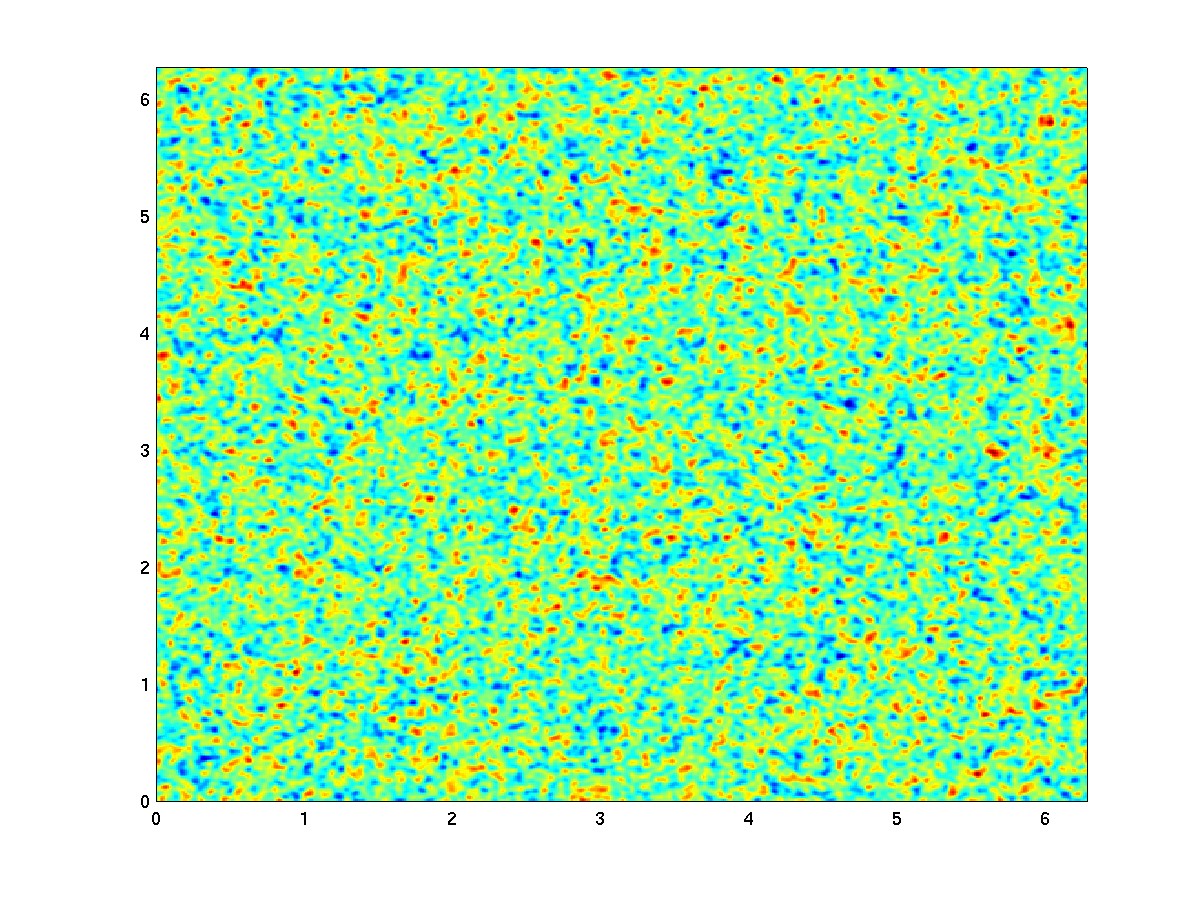}
\includegraphics[width=5cm,height=4.5cm]{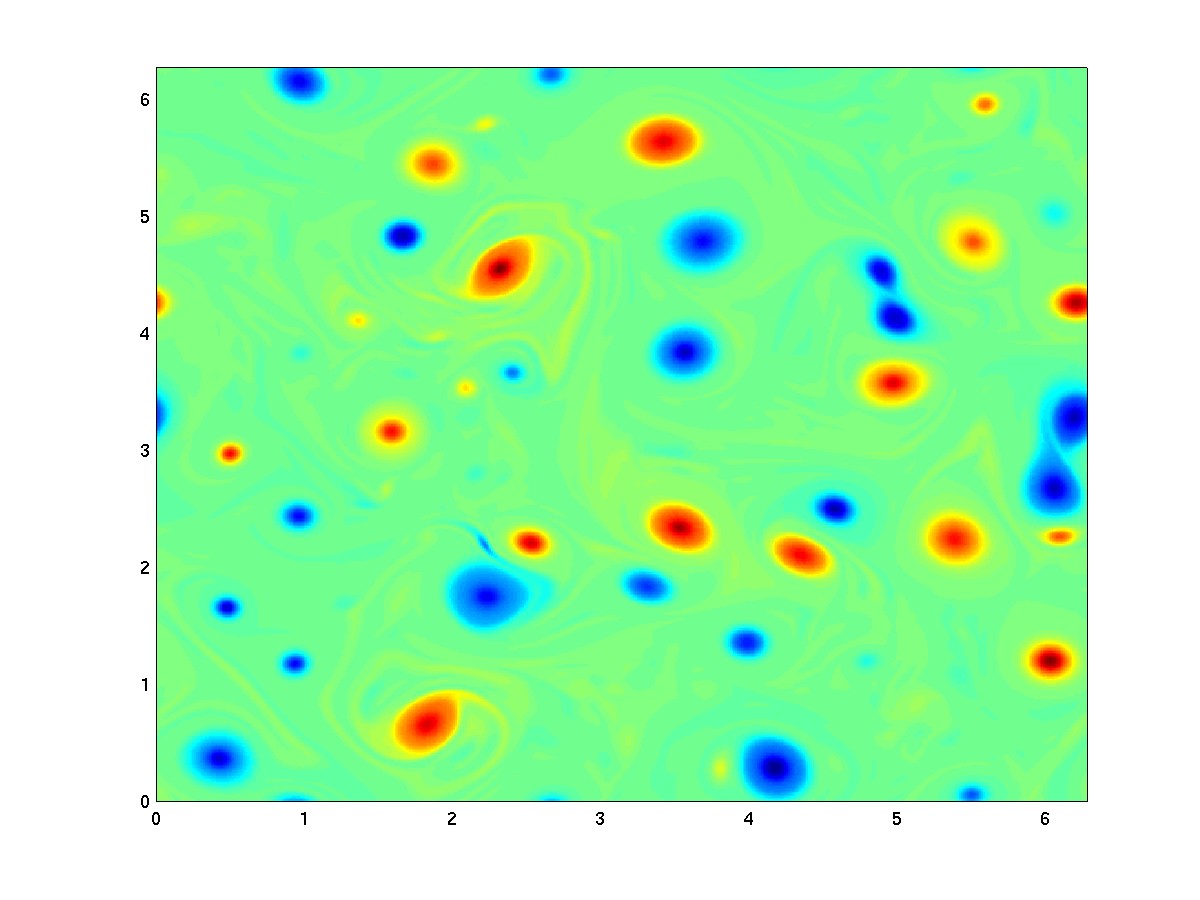}
\includegraphics[width=5cm,height=4.5cm]{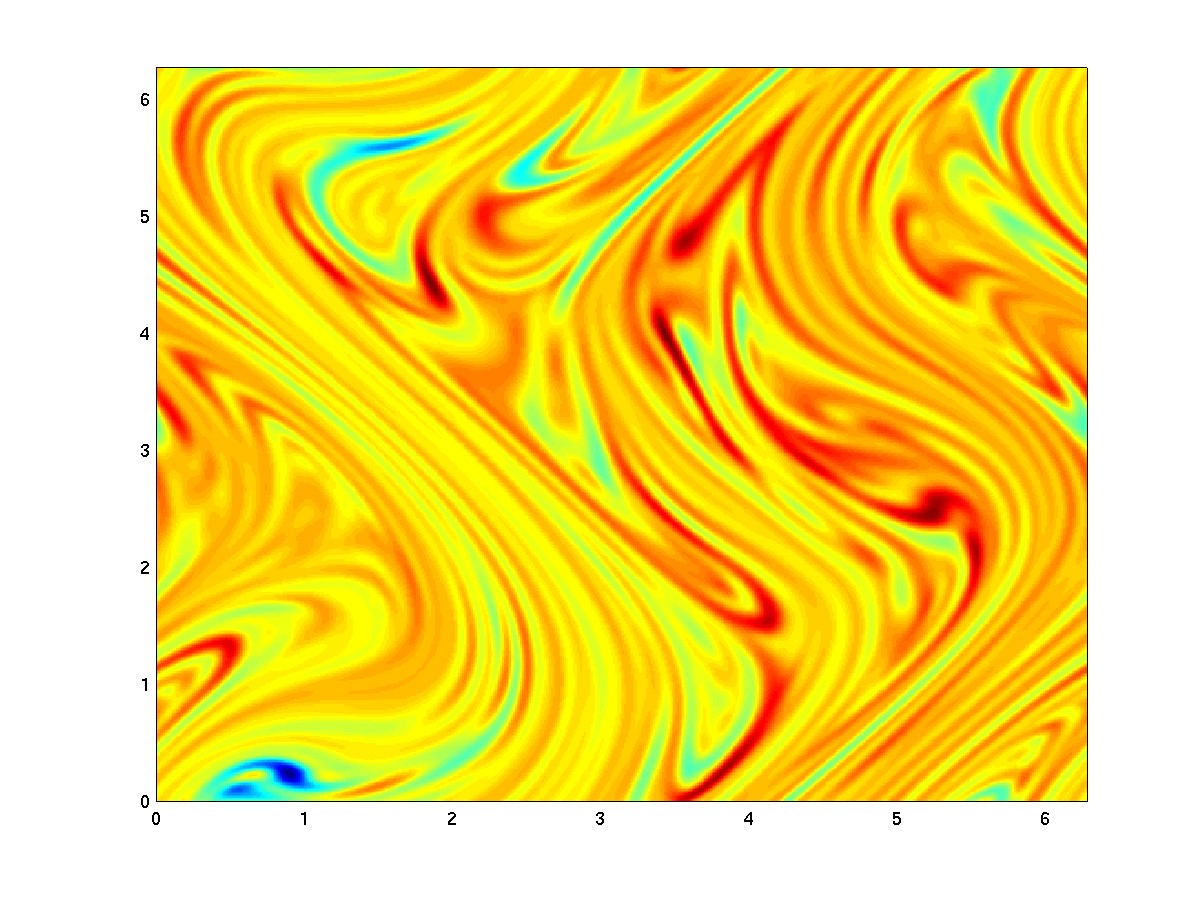}
\caption{\label{fig2} The first panel is the spatially un-correlated initial condition (smoothened 
via a diffusive stencil). The second and third panels show the emergent scalar field for 
$\alpha=0.5$ and $2$ respectively. Quite clearly, for $\alpha=0.5$ we have a field composed 
of coherent $\theta$ eddies while for $\alpha=2$ we obtain a filamentary geometry
reminiscent of a passive field when subjected to large-scale advection.}
\end{figure}

To develop a feel for the dependence of the geometry of an emergent scalar field on $\alpha$, we 
continue with non-dispersive simulations, though now from spatially un-correlated initial data 
chosen
from a Gaussian distribution with unit variance. Given the presence of an inverse
transfer of energy, we expect coherent structures to emerge from this un-correlated initial condition.
In fact, given the similar scales of the velocity and scalar fields for smaller $\alpha$ we
do not expect the scalar field to undergo much stretching and folding, while for large $\alpha$
we expect repeated events of this sort leading to a filamentary scalar field
--- much like the fate of a small-scale passive blob when advected by a large-scale smooth flow ---
due to the implicit large-scale strain (via the large scale-separation between $\theta$ and
$\psi$). These expectations are confirmed in Fig. (\ref{fig2}) which shows the initial condition and the 
emergent scalar field for $\alpha=0.5$ and $2$ respectively.

\section{The influence of the dispersive term}

\begin{figure}
\centering
\includegraphics[width=5cm,height=4.5cm]{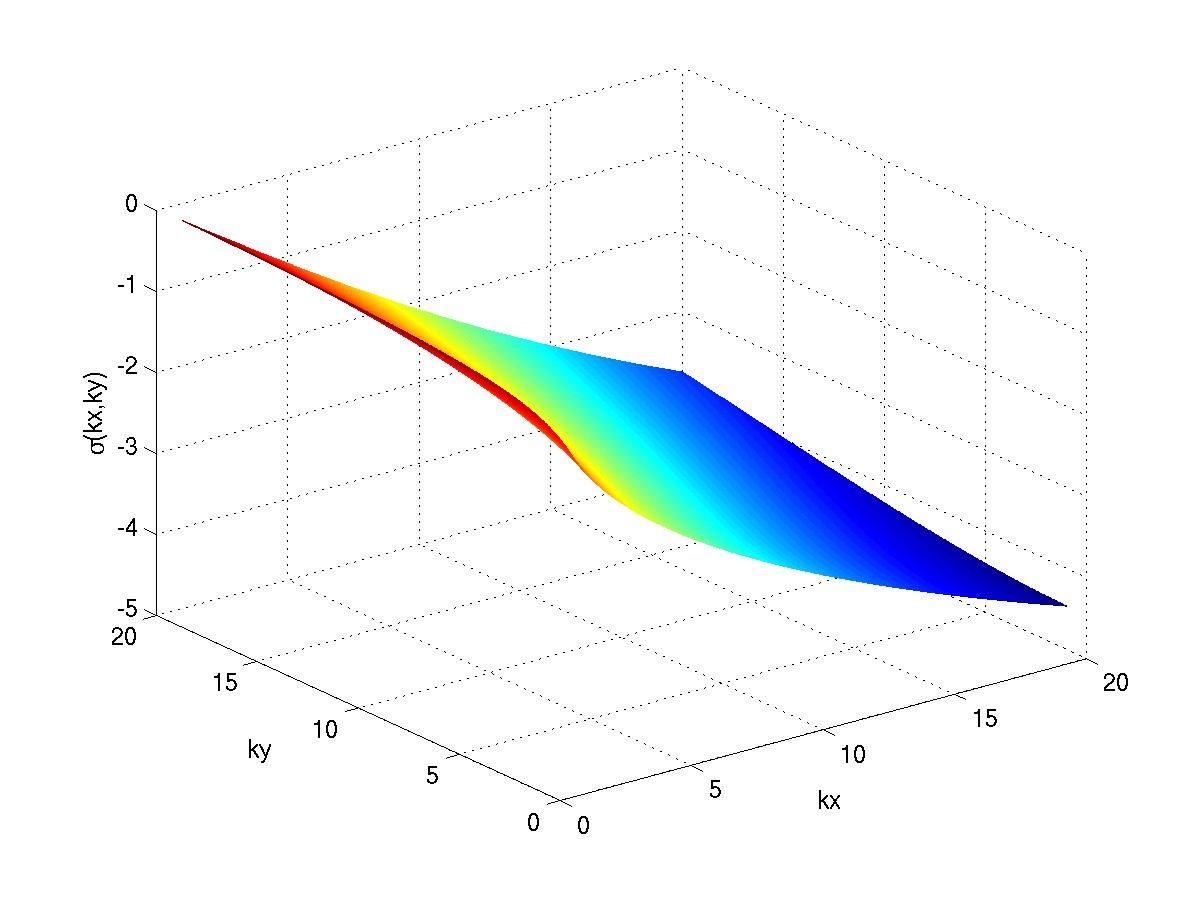}
\includegraphics[width=5cm,height=4.5cm]{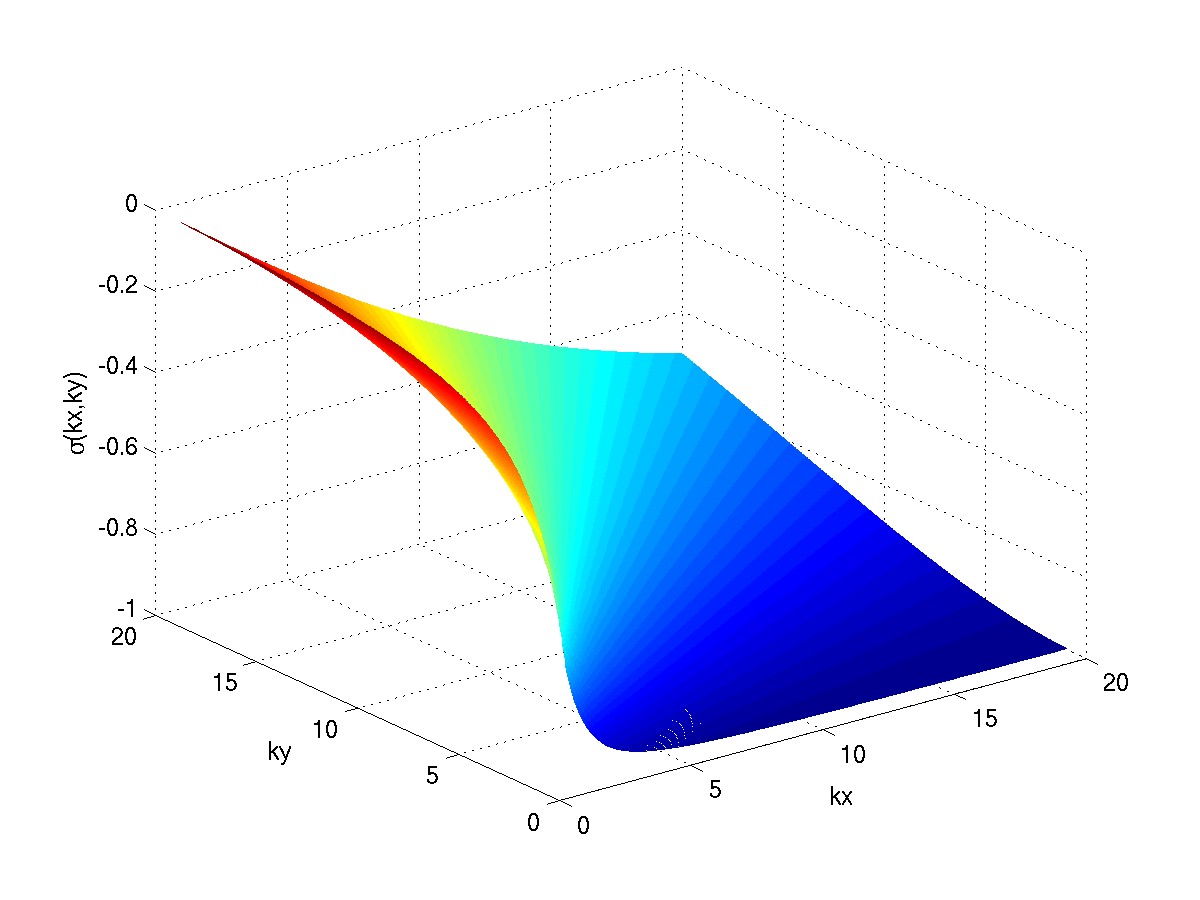}
\includegraphics[width=5cm,height=4.5cm]{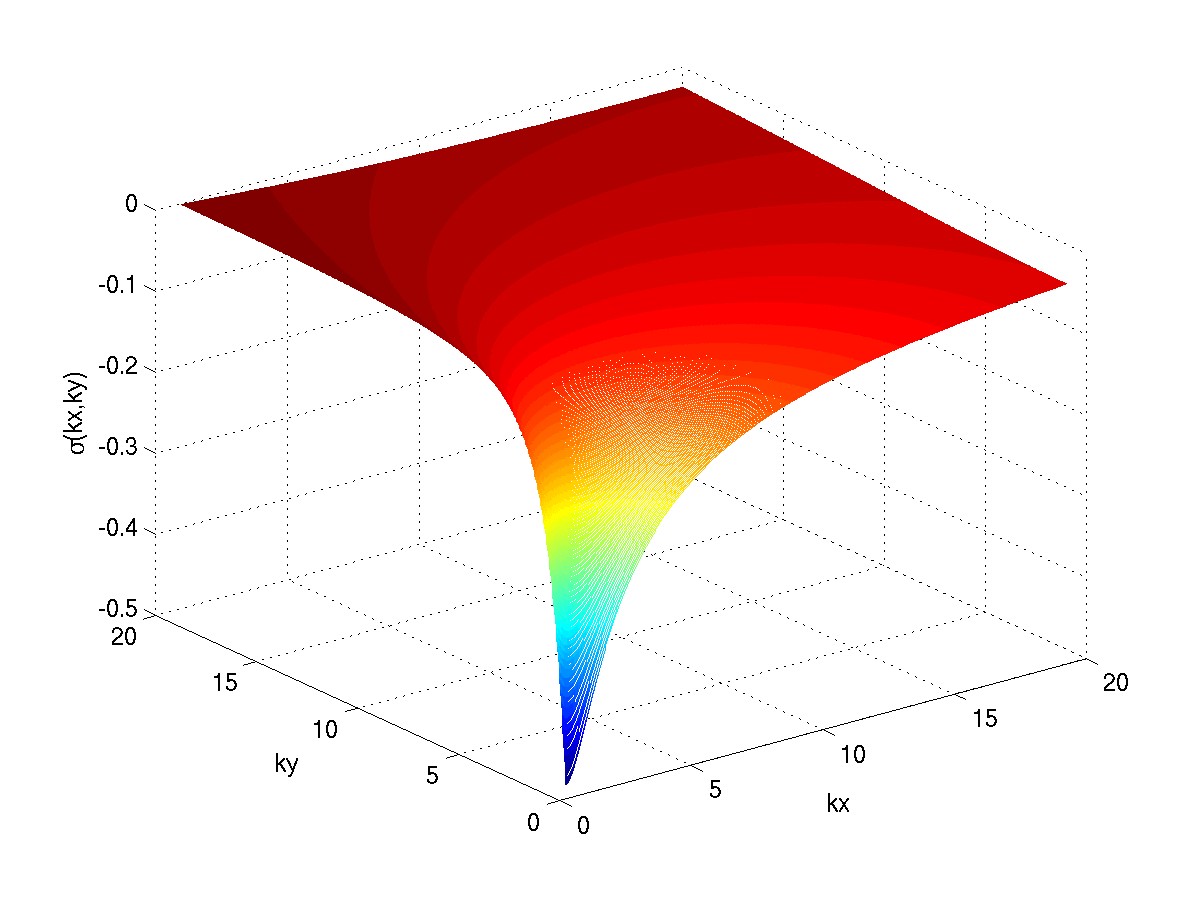}
\caption{\label{fig3} The dispersion relation (\ref{2}) with $\epsilon =1$ for, from left to right,
$\alpha=0.25,0.5$ and 1 respectively. Note that for $\alpha=0.5$ the frequencies are bounded while on either
side we obtain $|\omega| \rightarrow \infty$ in particular limits. }
\end{figure}

When the dispersive term is included, the linearized form of (\ref{1}) supports
waves with the dispersion relation

\begin{equation}
\omega(\vec{k}) = - \frac{k_x}{\epsilon k^{2\alpha}}~~;~~\textrm{where}~~ \vec{k} = (k_x,k_y)
\label{2}
\end{equation}
It is interesting to note that, for $\alpha=0.5$, (\ref{2}) gives $0 \le |\omega(\vec{k})| \le 1/\epsilon$.
For $\alpha < 0.5, |\omega| \rightarrow \infty$ for $k_y=0,k_x \rightarrow \infty$, whereas
for $\alpha > 0.5, |\omega| \rightarrow \infty$ for $k_y=0,k_x \rightarrow 0$. The dispersion 
relation for different $\alpha$ (on either side of $\alpha=0.5$) is plotted Fig. (\ref{fig3}). Quite 
clearly the frequencies have a very different dependence on wavenumber when $\alpha <,=$ or $> 0.5$. 
In fact, an important feature of the beta plane equations (which is true for all $\alpha > 0.5$) that 
$|\omega(\vec{k})|$ increases for large scales is no longer true for $\alpha \le 0.5$; in fact, 
for $\alpha < 0.5$ the smaller scale features have larger frequencies. \\

In the following, we occasionally refer to modes with zero frequency as slow modes and those
for which $\omega \neq 0$
as fast modes.
In a periodic setting, substituting a Fourier representation in (\ref{1}) yields

\begin{equation}
\frac{ \partial \hat{\psi}_k }{ \partial t} = \sum_{\mathcal T} C_{kpq} ~
\hat{\psi}_q \hat{\psi}_p  ~\exp\{ \frac{-{\rm i}}{\epsilon}[-\frac{p_x}{p^{2\alpha}} - \frac{q_x}{q^{2\alpha}} + \frac{k_x}{k^{2\alpha}}] {\rm t} \}
\label{3}
\end{equation}
where $\sum_{\mathcal T}$ represents a sum over $\vec{p},\vec{q}$ such that $\vec{k}=\vec{p}+\vec{q}$ and $C_{kpq}$
is the
interaction coefficient given by

\begin{equation}
C_{kpq}= (p_x q_y - p_y q_x)[ \frac{p^{2\alpha} - q^{2\alpha}}{k^{2\alpha}}]
\label{4}
\end{equation}

\subsection{Zonal flows}

Considering an initial value problem, the original deduction of anisotropic fields by Rhines \cite{Rhines}, in the
context of the beta plane equations was based
on the dual constraint of an upscale transfer of energy along with the tendency of
resonant triads to move energy into small frequencies (see
Hasselmann \cite{Hass} for a general demonstration irrespective of the details of the
nonlinear coupling co-efficient).
Indeed, the two pieces of the Rhines argument are : (i) energy moving to large scales as a result 
energy/enstrophy conservation, and (ii) the importance of resonant
triads in energy
transfer 
when dispersion modulates the advective nonlinearity.
In the context of the general dispersion relation
(\ref{2}),
for isotropic structures, $k_x \sim k_y \sim k$ we have

\begin{equation}
|\omega| = \frac{1}{\epsilon 2^{\alpha}} ~\frac{ 1 }{ k^{2\alpha -1} }.
\label{6}
\end{equation}
Hence for $\alpha > 0.5$, moving to large scales, i.e.\ for decreasing $k$ we encounter larger
frequencies. Therefore, to satisfy the dual constraints, Rhines \cite{Rhines} suggested that the system
would spontaneously generate anisotropic
structures; further examining (\ref{2}) shows that these constraints are satisfied for
$k_y \neq 0, k_x/k_y \ll 1$.
Also,
when considering energy transfer into large scales, i.e.\ $k < p, q$, interactions that fall
in the aforementioned anisotropic category are in fact near-resonant \cite{Salmon}.
In essence we have an anisotropic streamfunction that results in
predominantly zonal flows. Though note that for $\alpha < 0.5$,
decreasing $k$ implies smaller frequencies, therefore it is possible to maintain
isotropy while simultaneously transferring energy to large scales
and small
frequencies. Hence, for $\alpha < 0.5$, the dual constraints do not nessecitate the formation of 
a dominant of zonal
flow. Note that this does not imply zonal flows cannot form for $\alpha < 0.5$, in fact, 
substituting an expansion of the form $\psi=\psi^0 + \epsilon \psi^1...$, 
the $O(1/\epsilon)$ 
balance in (\ref{1}) yields $\partial \psi^0/ \partial x = 0 \Rightarrow u^0=u^0(y,t),v^0=0$. Of course,
this expansion doesn't imply any control over the higher order terms, but irrespective of $\alpha$,
at order zero, 
it indicates the possibility of zonal flow formation. \\

\begin{figure}
\centering
\includegraphics[width=8cm,height=8cm]{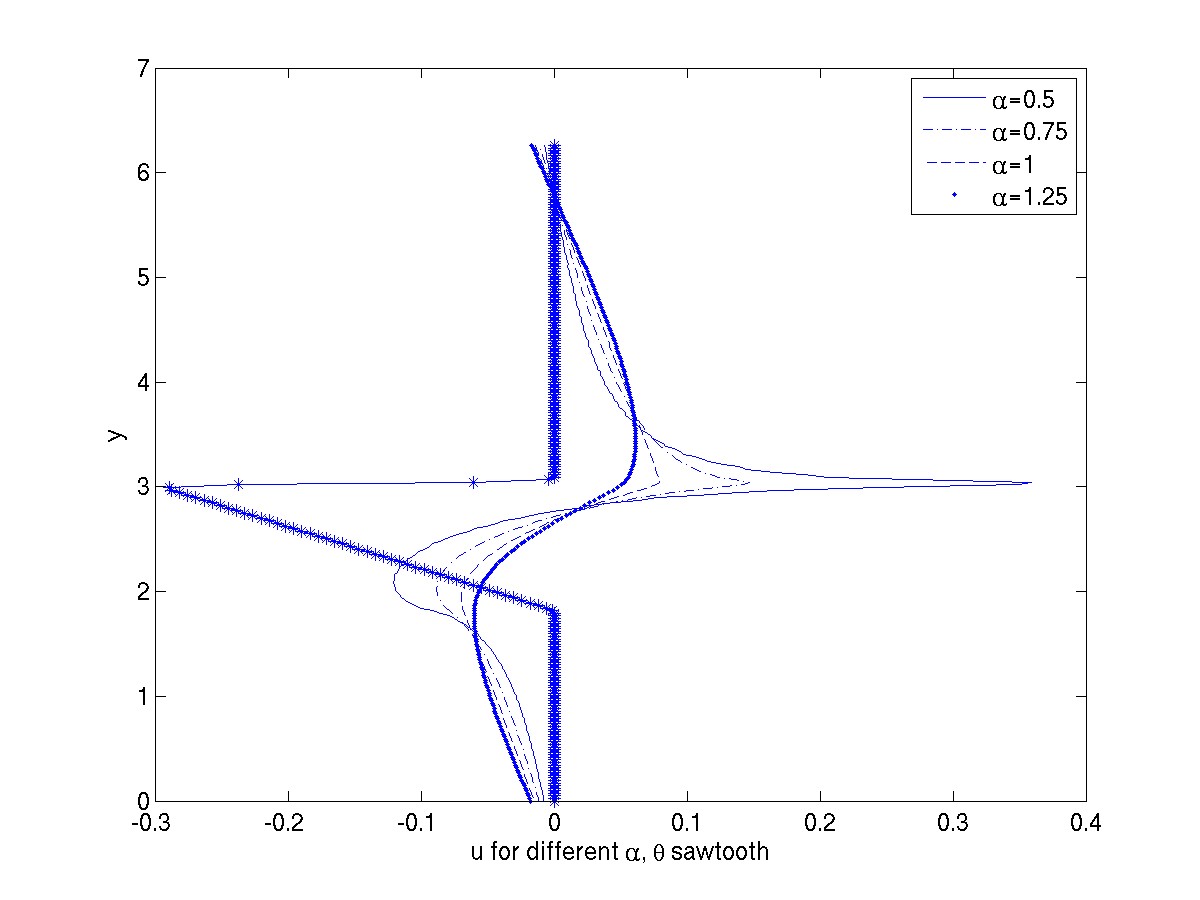}
\caption{\label{fig4} Inversion of a single "saw-tooth" $\theta$ profile for varying $\alpha$. As is evident,
in addition to the east-west asymmetry, smaller $\alpha$ (more local) gives stronger and narrower jets. }
\end{figure}

A different, though complementary, approach to the generation of zonal flows is to view (\ref{1}) as a
mixing problem, i.e.\ to consider the mixing of $\theta$ in the
presence of a background gradient \cite{Rhines-Young}.
In the context of the beta plane equations,
as shown by Rhines \& Young \cite{Rhines-Young},
steady flows with closed streamlines (or gyres)
result in the homogenization of
potential vorticity between these
streamlines.
With a linear background, this naturally leads to a "saw-tooth" vorticity
profile with high gradients concentrated on the streamlines \cite{McDrit}.
Further in a
statistically steady state, this reasoning
posits the existence of successive "saw-tooth" structures in the vorticity profile (or
in other words a potential vorticity "staircase") if
the size of the eddies is
smaller than the size of the domain \cite{McDrit} (see \cite{Danilov1}, \cite{Danilov2} for forced-
dissipative simulations and \cite{Marcus-Lee} for experimental results in this regard).
From this viewpoint localized zonal flows (jets)
result from an inversion of
the aforementioned potential vorticity "staircase". 
For the general family in (\ref{1}), $\theta + y/\epsilon$ is mixed by the flow and a similar 
homogenization within streamlines is expected to follow. Further, as the scale of the velocity field 
becomes smaller than that of the advected scalar when $\alpha$ crosses $0.5$ from above, we expect the homogenization
for small $\alpha$ to progress in a more diffusive manner, i.e.\ small-scale inhomogeneties will be erased
earlier which supports a gradual explusion of scalar gradients from small to large scales. In this vein,
Fig. (\ref{fig4}), much like the classic picture
by McIntyre, reproduced
in Dritschel \& McIntyre \cite{McDrit}, shows the inversion
of a single "saw-tooth" using the
general form in (\ref{1}).
As is evident --- apart from the known asymmetry in the east-west jets in the beta plane equations --- the more local
the inversion (i.e.\ for smaller $\alpha$) the narrower and
stronger are the corresponding jets. \\

\begin{figure}
\centering
\includegraphics[width=5cm,height=4cm]{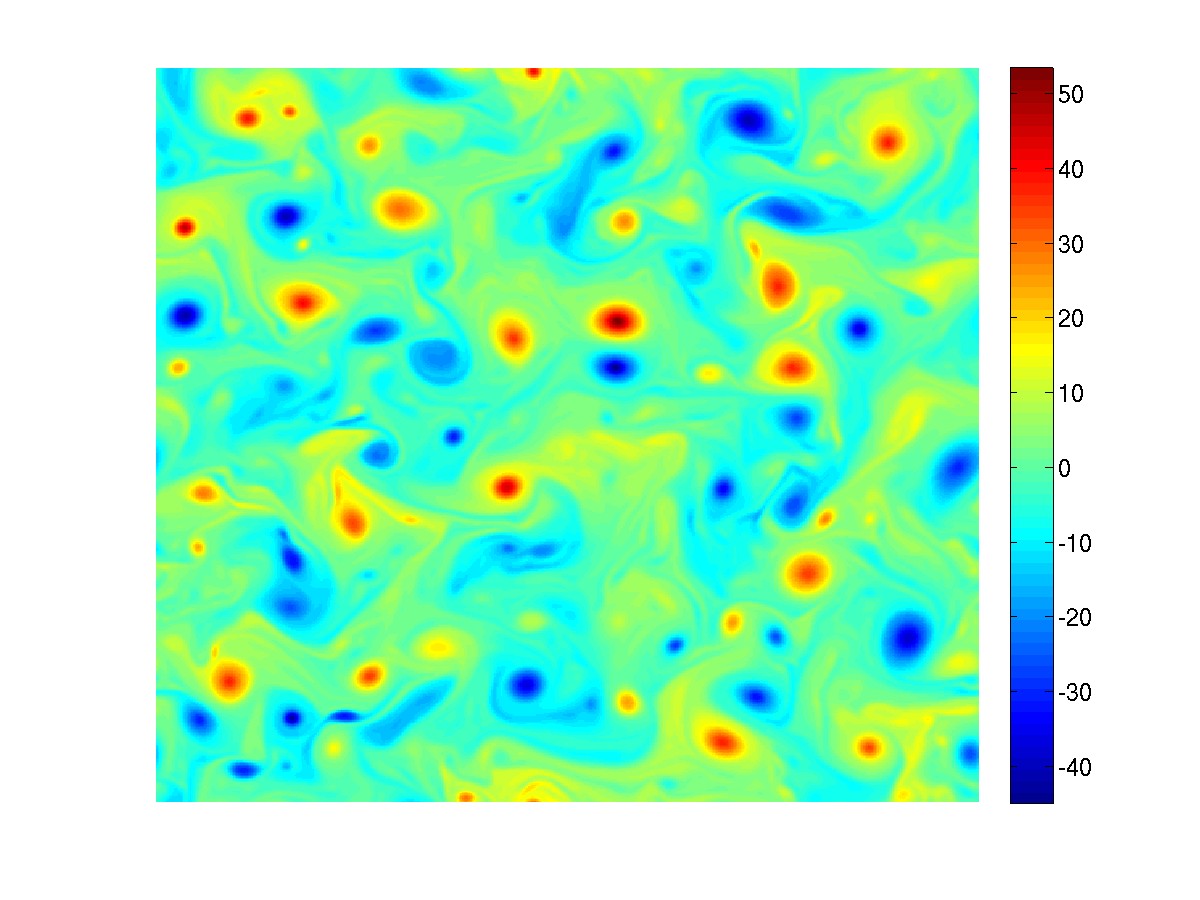}
\includegraphics[width=5cm,height=4cm]{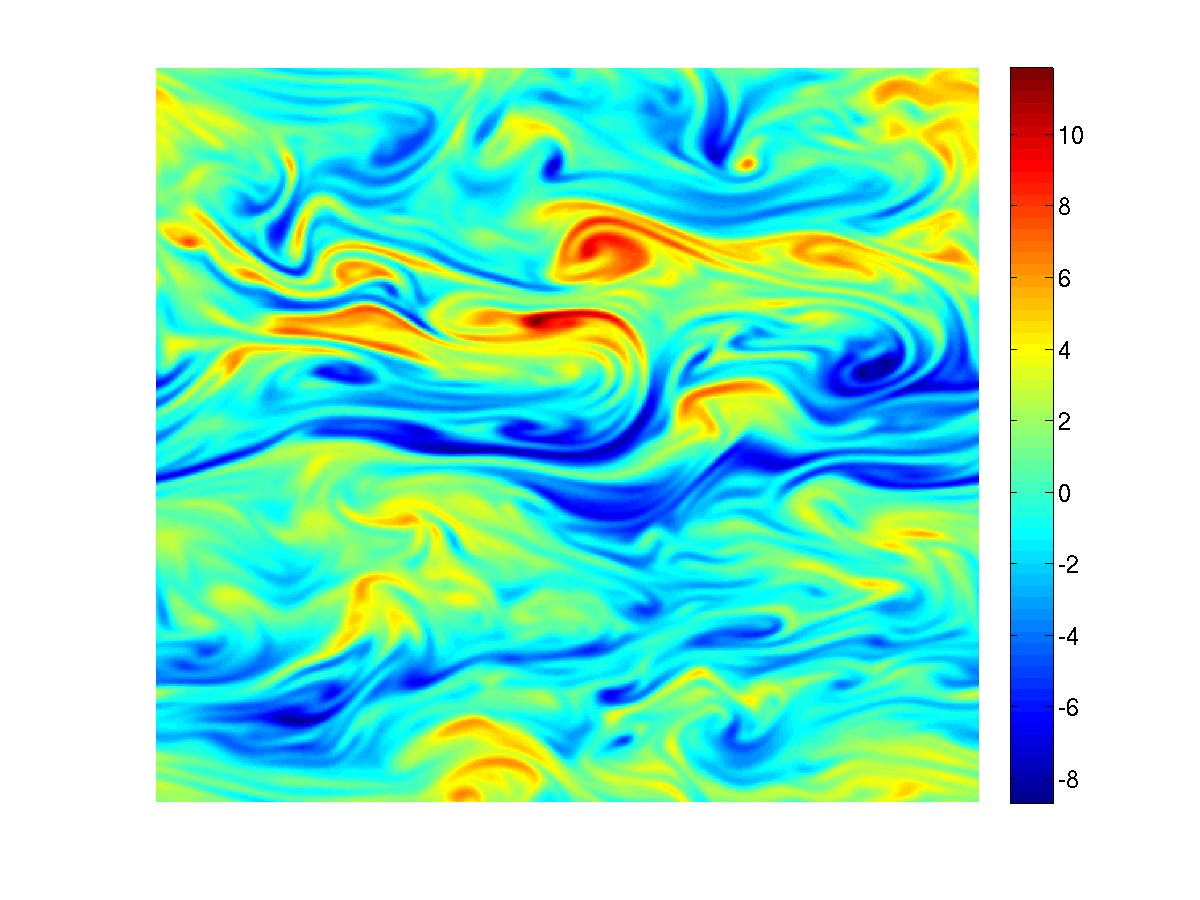}
\includegraphics[width=5cm,height=4cm]{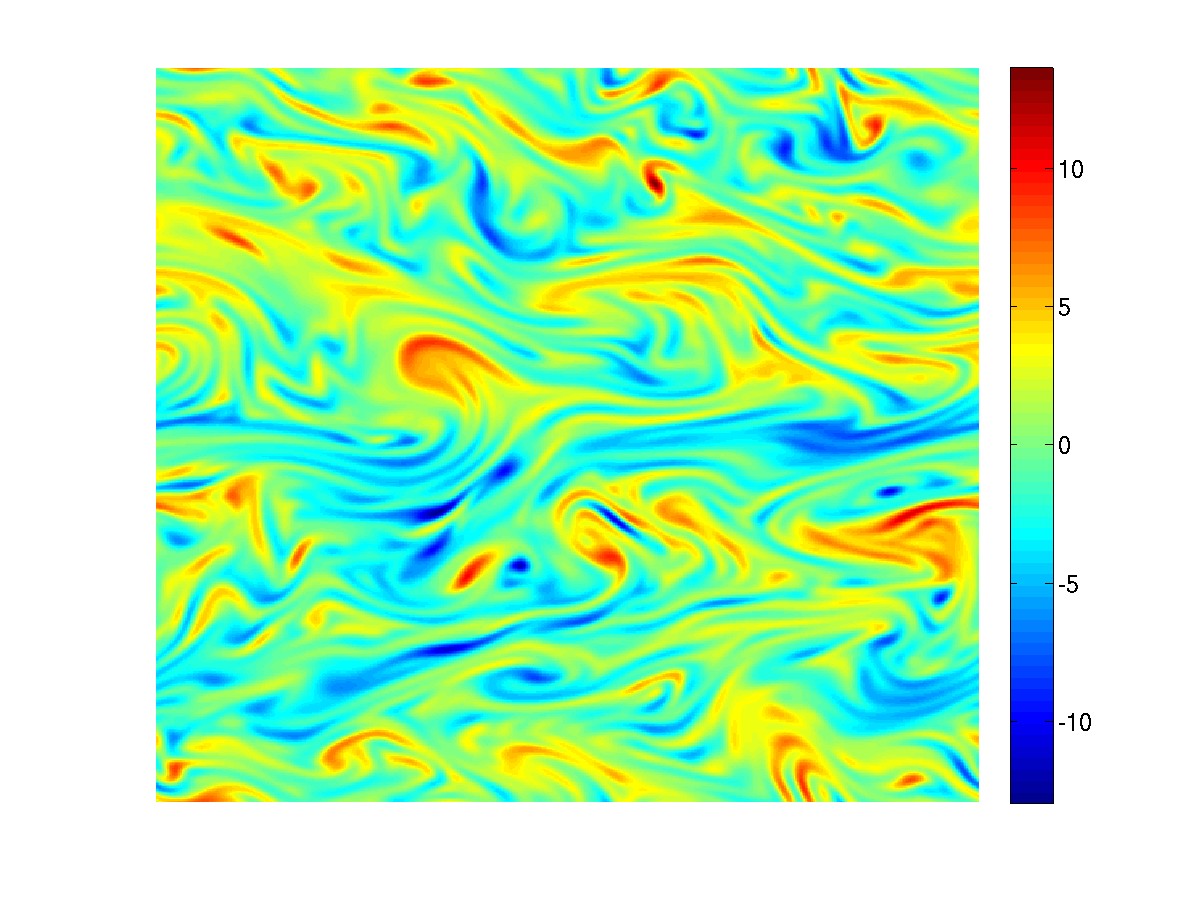}
\includegraphics[width=5cm,height=4cm]{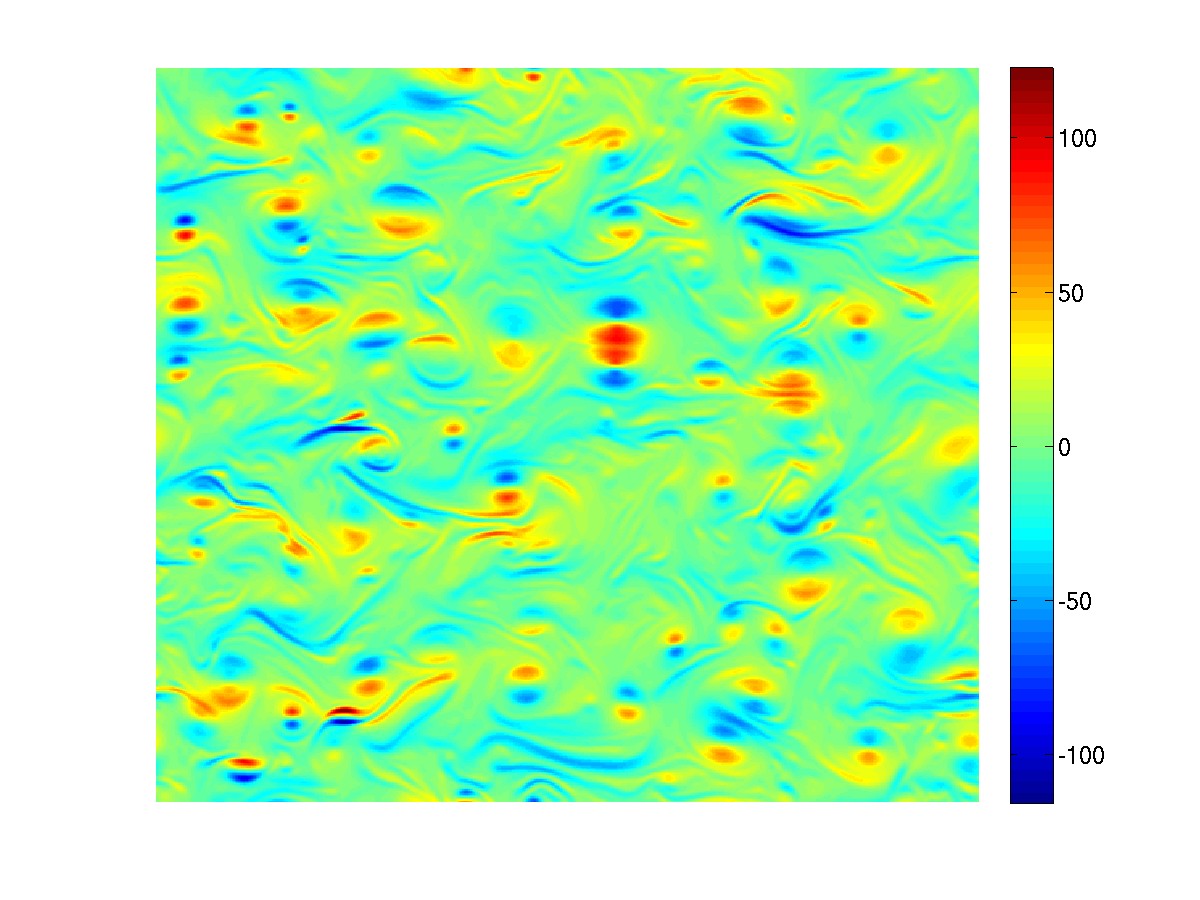}
\includegraphics[width=5cm,height=4cm]{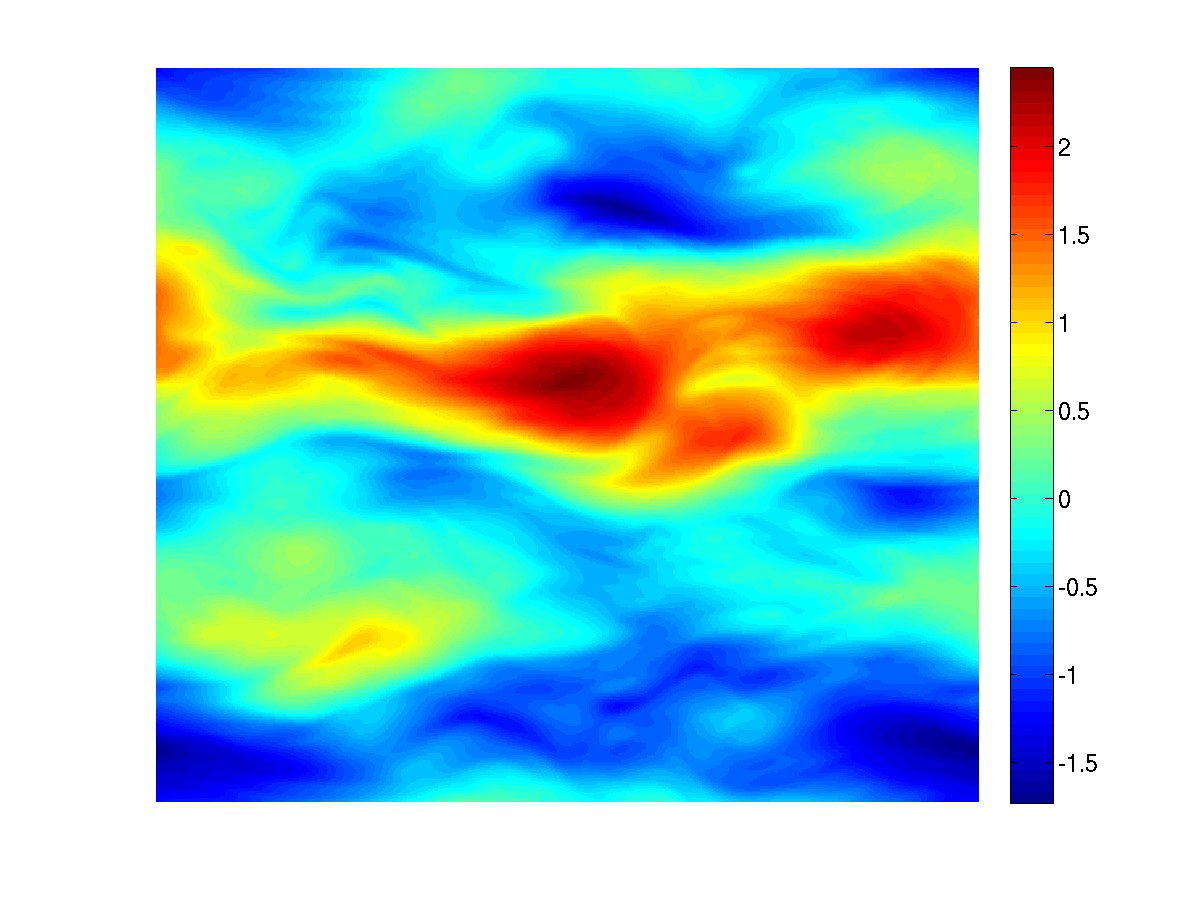}
\includegraphics[width=5cm,height=4cm]{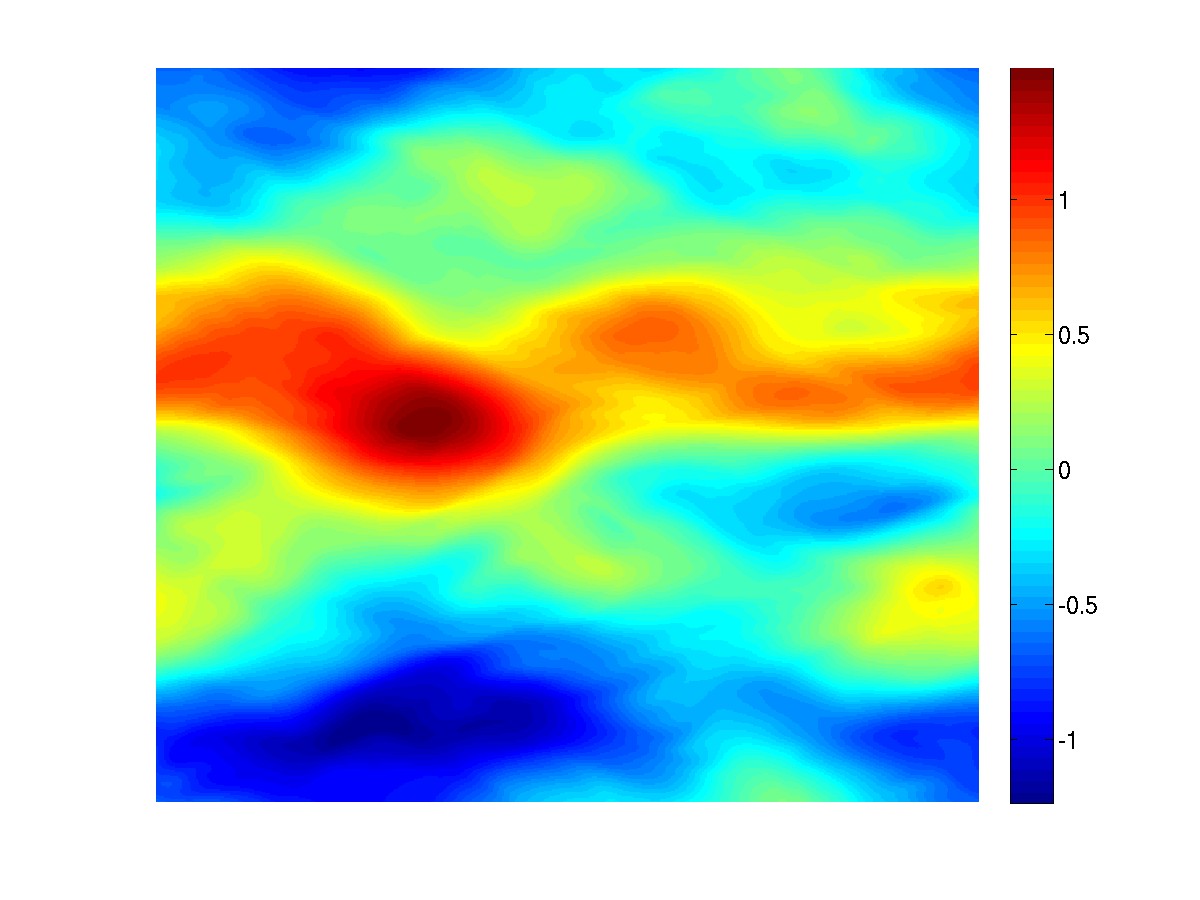}
\caption{\label{fig4a} $\theta, u$ fields
for $\alpha = 0.25,1$ and 1.25 in the upper and lower panels respectively. In all cases $\epsilon=0.1$.
Note the finer scale of the flow as compared to
the scalar field when $\alpha=0.25$. Further for increasing $\alpha$, we
obtain coherent zonal flows. }
\end{figure}

To examine the nature of emergent flows for differing locality in the presence of dispersion, we perform numerical simulations
with spatially un-correlated initial data as shown in the first panel of Fig. (\ref{fig2}). Setting
$\epsilon = 0.1$,
the scalar and zonal component of the velocity fields for $\alpha=0.25,1$ and $\alpha=1.25$ are shown in
Fig. (\ref{fig4a}), quite clearly
for $\alpha > 0.5$ we have, what might be termed a coherent
zonal flow. 
Note that, in accord with Fig. (\ref{fig4}), the flows are broader and
of smaller magnitude for increasing $\alpha$.

\subsection{Resonant and near resonant interactions}

Evolving (\ref{3}) in time, in the limit $\epsilon \rightarrow 0$, given the highly oscillatory 
nature of the integrals involved, 
the only terms that contribute to (\ref{3}) are those which are 
exactly resonant. 
Following the notation of Chen et\ al.\ \cite{Chen-etal}, i.e.\ the first entry in a triplet
$(\cdot,\cdot,\cdot)$ evolves via the interaction of the latter two entries, 
the limiting ($\epsilon \rightarrow 0$) dynamics are made up of :

\begin{itemize}
\item $p_x=q_x=k_x=0$ i.e.\ the slow-slow-slow ($s,s,s$) interactions : All such interactions between purely zonal flows are clearly resonant. However, from (\ref{4}) we
see that their interaction coefficients are zero. 

\item $k_x=0$, $p_x,q_x \neq 0$ i.e.\ slow-fast-fast $(s,f,f)$ or fast-slow-fast $(f,s,f)$ interactions :
It is easy to check that two modes with $\omega \neq 0$ cannot feed (extract) energy to (from)
one with $\omega=0$. In other words, 
two fast modes 
cannot generate a 
zonal flow. On the other hand, two fast modes can interact with each other using a 
zonal mode as a catalyst and this represents the passive advection of the fast modes via
a zonal shear flow.

\item $k_x,p_x,q_x \neq 0$ i.e.\ the fast-fast-fast $(f,f,f)$ interactions : These are the three-fast wave interactions. Though they are algebraically difficult to achieve,
these interactions represent fully 2D motion. 

\end{itemize}

As is evident, this situation is entirely analogous to the analysis of the beta plane equations as presented by 
Longuet-Higgins \& Gill \cite{LG}. 
Further, there is a close correspondence between  
these resonant interactions 
and the limiting dynamics 
(i.e.\ when the Rossby number $\rightarrow 0$) of 
rotating three-dimensional (3D) flows. 
For the 3D rotating 
case \cite{Babin-rot1}, the analogous set of $(s,s,s)$ interactions 
result in purely 2D flow (though now with a non-zero interaction coefficient); the 
second set of $(f,s,f)$ interactions leads to the passive advection of the 
vertical velocity via this 2D flow and inclusion of the third set of 
$(f,f,f)$
resonances yields the so-called $2\frac{1}{2}$ dimensional equations. 
Further, transfer from exclusively fast to slow modes
in pure rotation is also
precluded under resonance, i.e.\ 
$(s,f,f)$ interaction has a zero interaction coefficient, where the zero frequency mode 
corresponds to a columnar structure. 
Therefore, in the limit $\epsilon \rightarrow 0$ (Rossby number $\rightarrow 0$), 
it is not possible generate
a zonal flow (columnar structure) from exclusively non-zonal (non-columnar) initial data and
forcing. \\ 

Of course, the limit $\epsilon \rightarrow 0$ is usually not of great physical interest, in fact we are
usually more concerned with situations wherein $0 < \epsilon \le 1$. 
This opens the door to non-resonant
interactions, and quite naturally, we expect near-resonances --- i.e.\ triads for which
$\omega(\vec{p})+\omega(\vec{q})-\omega(\vec{k}) = \delta \ll \epsilon$ --- to be of greater
significance than the remainder that are, in terms of ($\delta,\epsilon)$, far from resonance.
As the $(s,f,f)$ non-resonant triads have a non-zero interaction co-efficient
we have the possibility of directly transferring energy from non-zonal to
zonal modes. 
Choosing $\delta$ for given $\epsilon$, such transfer to slow modes
has been numerically verified
in the beta plane and 3D rotating systems \cite{Smith-qg}, \cite{Smith-rot}. 

\section{The inverse and forward cascades}

With the aim of examining a systematic transfer of energy/enstrophy from one scale 
to another, we proceed to numerical simulations from weakly non-local to local conditions (i.e.\
$\alpha \le 1$) \cite{Kraichnan1967}, \cite{Chen-2D}.
The numerical scheme is a de-aliased psuedospectral
method with fourth order Runge-Kutta time stepping.
All simulations are performed in Matlab at a resolution of $750 \times 750$.
Further, we supplement the RHS of (\ref{1}) with
hyper-viscous dissipation
of the form $(-1)^{n+1} \nu_n \triangle^{n} \theta$ (to act as a sink at small scales,
with $n=4$). As per Maltrud \& Vallis \cite{Maltrud-Vallis} we choose
$\nu_n = \theta_{\textrm{rms}}/({k_m}^{2n-2})$, where
$k_m$ is the maximum resolved wavenumber. \\

Denoting the domain, forcing and dissipation
scales by $k_L, k_f$ and $k_d$ respectively, we consider $k_L \ll k_f < k_d$ ($k_L < k_f \ll k_d$) which
allows us achieve, for a given resolution, as large an inverse (forward) energy (enstrophy) transfer
regime as is possible.
The forcing is of random phase at every time-step and de-correlates with a time-scale $\tau$ (this 
is chosen to be comparable to the fastest linear wave in the model).
Further the forcing is localized to a few wavenumbers, and
it is important to note --- especially in the context of a small-scale case ---
that our forcing is not
chosen in a "ring" of waveneumbers
(i.e.\ it is not localized by means of $\exp\{ -(|\vec{k}| - k_f)^2 \}$)
but rather,
we strictly ensure that there is no projection of the forcing on small $k_x$ or $k_y$ (i.e.\
the localization is by means of $\exp\{ -(k_x - k_f)^2 - (k_y - k_f)^2 \}$). This ensures that if we see
the formation of zonal flows, it is due to the transfer of energy from modes that are,
in a sense, "far from zonal" to zonal
and near-zonal modes \cite{foot0}. Finally, we also include a 
linear damping on the largest scales,
of the form $-\lambda \theta$, which allows us to achieve a stationary state. In some of the 
inverse transfer simulations the time required to achieve stationarity is prohibitively large. In 
such cases the
runs are halted when the power-law scaling ceases to evolve (i.e.\ only the very largest scales are still growing).

\subsection{Inverse transfer : Small-scale forcing}

As mentioned, in the context of the beta plane equations,
one observes the spontaneous generation of zonal flows from isotropic
random small-scale forcing (recent numerical simulations can be found in \cite{Danilov1}, \cite{Danilov2}).
In line with Rhines' argument for a non-isotropic streamfunction and dominant zonal flow,
the reasoning --- supported by numerical simulations \cite{Maltrud-Vallis}, \cite{Vallis-Maltrud} --- is that,
the upscale transfer of energy is isotropic up to the so-called
Rhines Scale, beyond which energy is preferentially deposited anisotropically in wavenumbers
that satisfy $k_x/k_y \ll 1$
\cite{Hol-gafd},\cite{Salmon}. In terms of the scaling,
some forced simulations employing both large and small-scale sinks support the dimensionally predicted
isotropic $k^{-5/3}$ and anisotropic ${k}_y^{-5}$ power-laws for the non-zonal and zonal energy spectra
respectively \cite{Cheklov-etal}, \cite{Sukoriansky-etal}. Our aim is to examine the 
inverse transfer of energy for different $\alpha$, and its sensitivity, for a given $\alpha$, 
to the strength of the the dispersive
term. Therefore we do not focus on the emergent zonal flows, but rather on the scaling 
associated with the inertial-range \cite{foot-in} 
under weakly-local to local energy transfer. \\
 
\begin{figure}
\centering
\includegraphics[width=8cm,height=7.5cm]{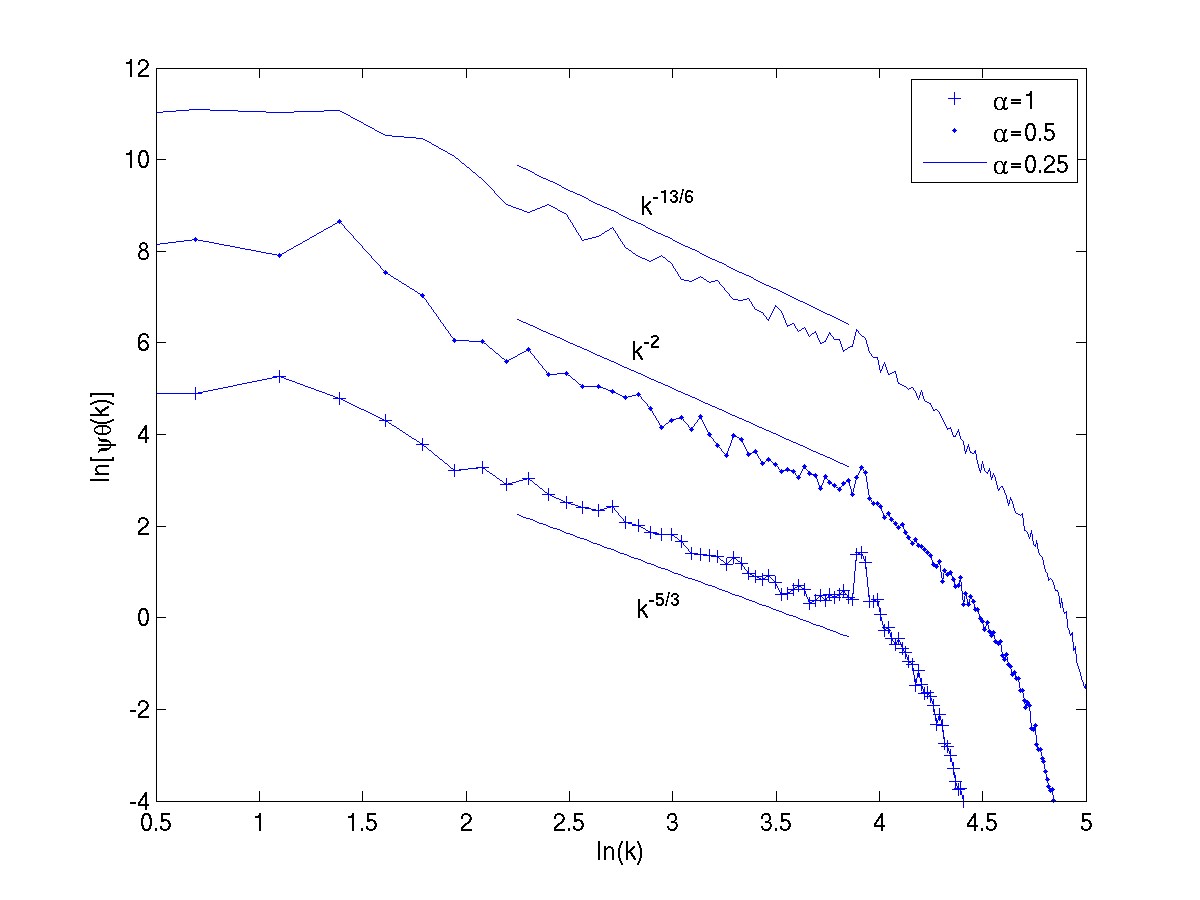}
\includegraphics[width=8cm,height=7.5cm]{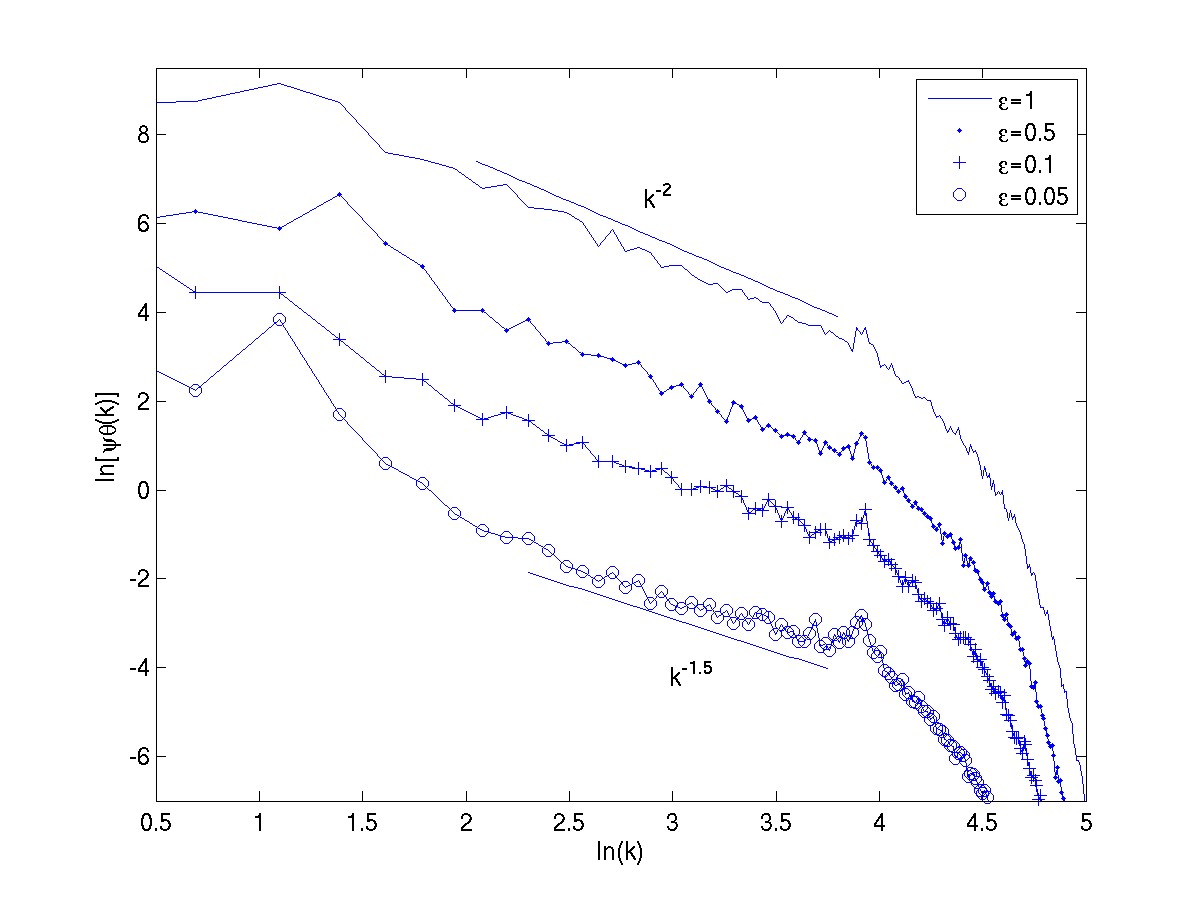}
\caption{\label{fig5} The first panel shows energy spectra in the weakly-local to local inverse energy 
transfer regime 
with $\epsilon=0.5$. 
We notice well-developed power-laws, and the slopes are 
in reasonable agreement with non-dispersive estimates. Note that these are slopes from individual realizations, not
temporal or ensemble averages. The second panel shows the inverse transfer for $\alpha=0.5$, 
with $\epsilon=0.05,0.1,0.5$ and $1$. The sensitivity of the scaling with respect to $\epsilon$ is evident,
also note that the range up to which the scaling extends decreases with $\epsilon$.}
\end{figure}

Utilizing small-scale random forcing and large-scale damping as prescribed in the previous section, 
the energy spectra for different $\alpha$ with $\epsilon=0.5$ are shown in Fig. (\ref{fig5}). 
The non-dispersive estimate 
for an inverse cascade reads
$E(k) \sim k^{-(7-2\alpha)/3}$ \cite{PHS}; 
quite clearly we obtain power-laws that are in reasonable agreement with the aforementioned dimensional estimate.
Of course, our resolution is moderate and hence we do not pursue absolute quantitative accuracy in 
terms of ensemble simulations with appropriate error estimates.
We now repeat these experiments with a dispersive term of differing strength ($\epsilon=1,0.1,0.05$). 
The results for $\alpha = 0.5$ are shown 
in the second panel of Fig. (\ref{fig5}), as is evident, along with a shorter inertial-range 
the slopes 
become shallower with a progressively stronger dispersive term. 
The shortening of the inertial-range is expected as, it is easy to verify that 
the Rhines scale decreases with $\epsilon$. At present we do not have an explanation for the 
decrease in the slope of the power-law. Indeed, as the set of near-resonant
interactions becomes smaller with decreasing $\epsilon$, it becomes prohibitively
expensive to address these interactions in an adequate (or, at least, numerically consistent) manner.
Hence, 
at our present resolution of $750 \times 750$ we can only claim an 
agreement with non-dispersive estimates to be valid for $\epsilon \sim 1$. We also note the 
more pronounced steepening of the spectrum beyond the inertial-range for smaller 
$\epsilon$, 
though clearly we do not have adequate 
resolution to claim any spectral form for this steepening.

\subsection{Forward transfer : Large-scale forcing}

In the beta plane equations, the forward cascade of enstrophy has been studied by Maltrud \& Vallis
\cite{Maltrud-Vallis}. It was noticed that the energy spectrum had non-universal properties,
in particular they observed power-laws of the form $k^{-\sigma}$ with $\sigma \in [3,3.5]$ 
\cite{Maltrud-Vallis}. 
From (\ref{3}), it is evident that the
dispersive term reduces the "effectiveness" of the nonlinearity; a conjectured consequence of which is the
steepening of the energy spectrum from its dispersionless $k^{-3}$ form \cite{foot2}.
Adopting a mixing perspective, the evolution of a large-scale initial condition (and its comparison
with an identical passive field) in the beta plane equations was considered by
Pierrehumbert \cite{Ray-gafd}. Indeed, the
rapid emergence of fine-scale features in the $\theta$ field (implying a forward transfer of $\theta$ variance)
was seen to
yield a relatively high-wavenumber $k^{-1}$ enstrophy spectrum. Much like our earlier discussion,
the reasoning revolved around the passive driving of the high-wavenumber (or small-scale)
features of the $\theta$ field. Indeed, the $k^{-1}$ spectrum coincides with the scaling of a
small-scale passive field (in the convective range) driven by means of a large-scale
smooth advecting flow, i.e.\
the Batchelor regime \cite{Batchelor1959},\cite{Kraichnan1974}. It is quite interesting that,
the small-scale features generated from the initial large-scale $\theta$ field do not
spontaneously "re-combine" to, once again, result in the development of large-scale structures.
In fact, the absence of any smooth, persistent coherent structures is also evident in the
$k^{-1}$ scaling, i.e.\ via the lack of anticipated
spectral steepening that results from the presence such features. \\

\begin{figure}
\centering
\includegraphics[width=8cm,height=7.5cm]{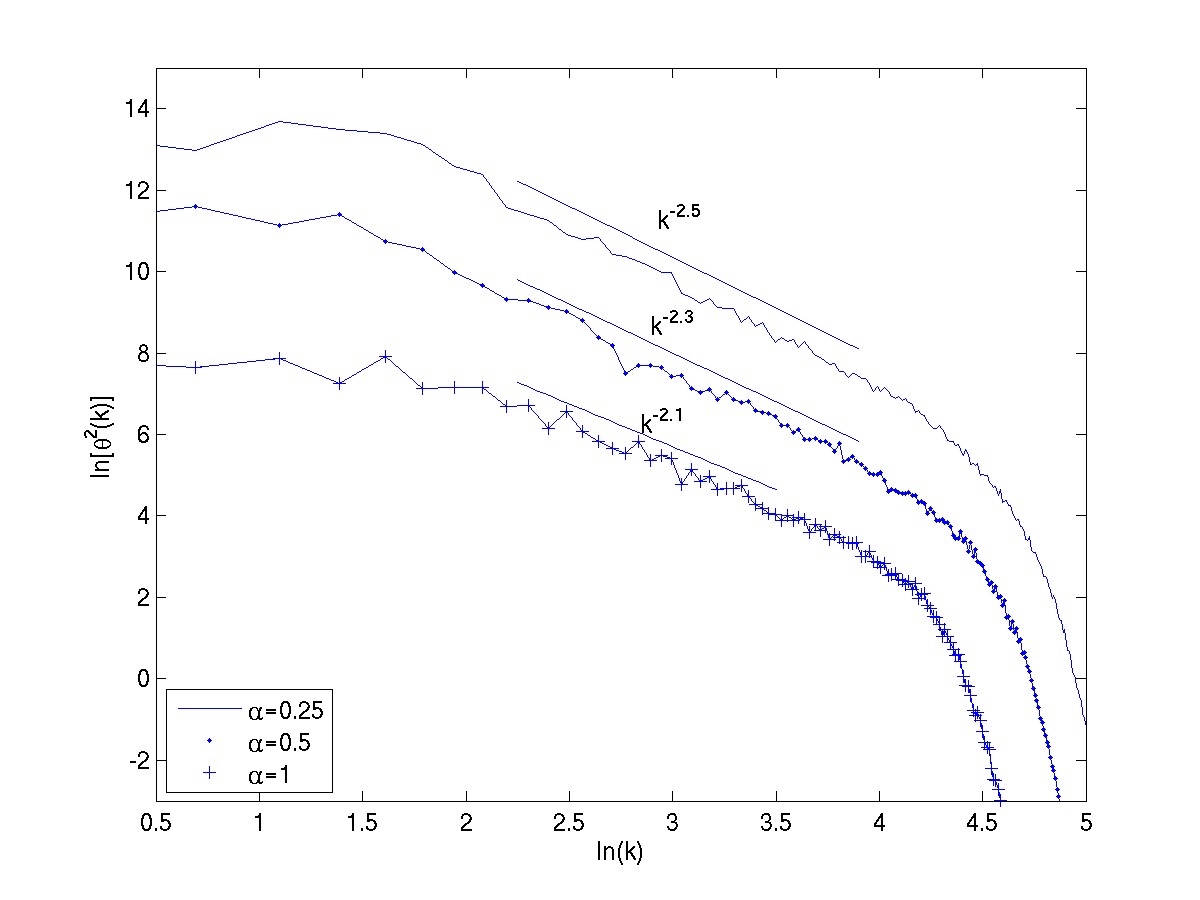}
\includegraphics[width=8cm,height=7.5cm]{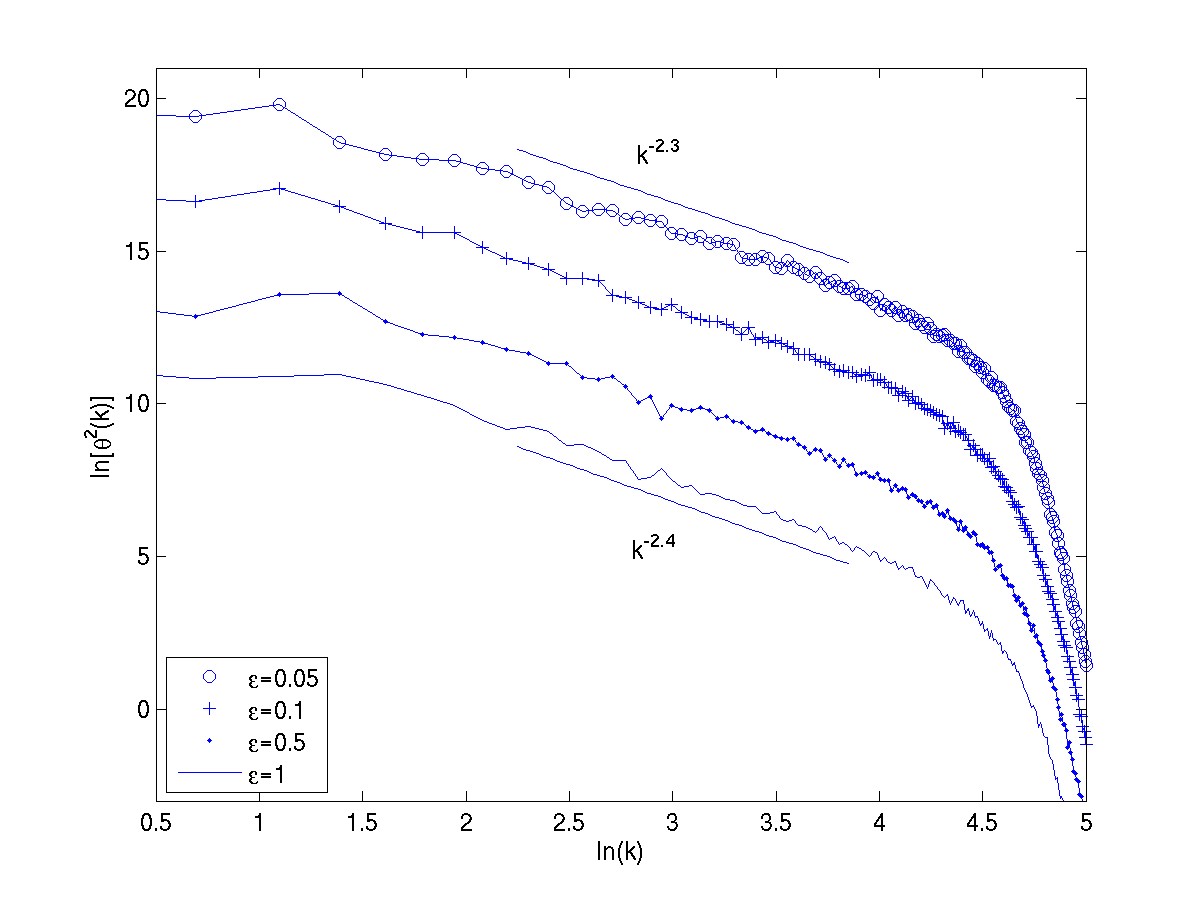}
\caption{\label{fig6} Enstrophy spectra in the forward weakly-local to local enstrophy transfer regime
for $\epsilon=0.5$. 
We notice well-developed power-laws, but the slopes are 
significantly steeper than non-dispersive estimates. The second panel shows the same experiment for $\alpha=0.5$
but with varying $\epsilon$ ($\epsilon=0.05,0.1,0.5$ and $1$ from top to bottom). 
Quite clearly, the slopes are fairly insensitive to
$\epsilon$. Once again, we note that our estimates of the slopes are based on invidual realizations as 
opposed to a temporal or ensemble average that might be more suitable at higher resolutions for 
accurate quantitive slopes and error estimates. }
\end{figure}

In addition to
the increased physical locality seen in Fig. (\ref{fig0}), from a spectral perspective
the forward cascade is local for $\alpha < 1$ \cite{PHS}, \cite{Sch},
($\alpha=1$ is the well-known logarithmically divergent
case \cite{Kraichnan1967}, \cite{Falk-Leb}).
In fact, the increased contribution from local triads
(i.e.\ all legs of the triads are of comparable size) 
to the forward cascade has been noted in recent simulations comparing the
SQG to the 2D Navier-Stokes case \cite{Watanabe2}. 
Possible consequences of the increased locality, such as the 
development of frontal discontinuities, the fractal nature of iso-$\theta$ level sets and the 
multifractal nature of the dissipation field have been examined in a series of SQG 
simulations \cite{Oki}, \cite{Jai-sqg}. 
As per non-dispersive estimates, the enstrophy
spectra, shown in Fig. (\ref{fig6}), have steeper slopes for increasing locality. 
Though, unlike the previous estimates for the inverse energy cascade, the slopes 
are clearly much steeper than the non-dispersive similarity hypothesis, which for a forward 
enstrophy cascade reads ${\mathcal E}(k) \sim k^{-(7-4\alpha)/3}$ 
\cite{PHS}. Repeating the experiments for $\epsilon=1,0.1,0.05$ we always see the development 
of a clear inertial range; though now, as seen in the second panel of Fig. (\ref{fig6}) 
which shows the $\alpha = 0.5$ case,
the slopes are insensitive to $\epsilon$. 
Indeed, the principal difference is in the time 
required for a stationary state to emerge. Specifically, as the strength of the dispersive term 
increases (smaller $\epsilon$) the simulations have to be carried out for progressively longer durations
(though, it is worth noting that these forward transfer runs are much faster than the previous
set of simulations involving inverse energy transfer).

\section{Conclusions and discussion}

We considered the evolution of a family of dispersive dynamically active scalar fields, the 
members of this family are characterized by the degree of locality in the $\psi-\theta$ relation.
Further, the dispersive term is strong, i.e.\ it is preceded by the inverse of a small parameter 
$0 < \epsilon \le 1$.
With regard to the effect of $\alpha$ on the energy/enstrophy transfer, according to the Fjortoft estimate, 
there is a monotonic relation between increasing $\alpha$ and 
the fraction of energy (enstrophy) transferred to large (small) scales. Further, the bias in energy/enstrophy
partitioning increases with the spectral nonlocality of the transfer.
As for the geometry of the scalar field, large isotropic coherent eddies are seen to develop from spatially un-correlated 
initial data for small $\alpha$, while for large $\alpha$ we 
see the formation of a filamentary structure (much like a small-scale passive field when advected via a large-scale 
smooth flow). 
Rhines' argument for the spontaneous 
asymmetry in the streamfunction is seen to hold for all $\alpha > 0.5$, while for $\alpha < 0.5$
it is possible to maintain isotropy while satisfying the dual constraints of energy transfer to 
large scales and small frequencies. Indeed, simulations from un-correlated initial data confirm
the emergence (and lack thereof) of dominant coherent zonal flows 
for $\alpha > 0.5$ ($\alpha < 0.5$)
respectively \cite{footer}. \\ 

Utilizing random small-scale forcing, 
for weakly non-local to local conditions (i.e.\ for 
$\alpha \le 1$), we observe a clear power-law scaling in the energy spectrum, and for 
$\epsilon \sim 1$ the slope of the
spectra agree with non-dispersive estimates. 
Under large-scale forcing, much like the inverse energy cascade, 
we see a clear forward 
cascade of enstrophy accompanied by a power-law scaling; though in contrast to the inverse regime, 
the slopes are much steeper than
non-dispersive similarity hypotheses. 
Repeating the experiments with a dispersive term of differing strength shows that the 
scaling depends on $\epsilon$ in the inverse transfer regime. Specifically, along 
with an expected shortening of the inertial-range, at our present resolution, the 
associated slopes also 
become progressively shallower as the dispersive terms becomes stronger.
On the other hand, the forward transfer regime is quite insensitive to the strength of the 
dispersive term, in fact we observe the enstrophy cascade slopes to be fairly universal 
with respect to the range of $\epsilon$ values considered. \\

With regard to the mathematical aspects of (\ref{1}), $\alpha=0.5$ is known to be
special in the sense that it
represents an open problem with regard to
global regularity of non-dissipative solutions (see for example Constantin, Majda \& Tabak \cite{Const-NL}
for an analogy between front formation in SQG and finite time
singularities in the 3D Euler equations).
In fact, present estimates for well behaved solutions
require dissipation of the form $\triangle^{\rho}$ with $\rho=0.5$ for both, the non-dispersive and
dispersive cases \cite{KNV},\cite{Kiselev}. Unfortunately,
regularity results for $0.5 < \alpha < 1$ are presently un-settled. It is unclear
if the difficulty in
achieving regularity (by present techniques) arises abruptly at $\alpha=0.5$ or whether one
requires gradually stronger dissipation as
$\alpha$ decreases from unity. Physically, this is interesting as $\alpha=0.5$ is precisely the 
border at which the advecting flow is of the same scale as the advected field. In fact,
in the context of both the dispersive and non-dispersive cases, it would be quite interesting to know if 
a change in relative scales of the advecting and advected fields (as $\alpha$ crosses 0.5 from above) 
is connected to the deeper mathematical issue of 
the regularity of solutions. \\

{\it Acknowledgements : }
We would like to acknowledge discussions with Prof.\ A.J. Majda (Courant Institute, NYU) and  
Prof.\ A. Kiselev (Math Department, Wisconsin, Madison). 
We would also like to thank the reviewers for their extensive comments.
Financial support 
was provided by NSF
CMG 0529596 and
the DOE Multiscale Mathematics program (DE-FG02-05ER25703). J.S.\ would also like to acknowledge the hospitality 
of the Indian Institute of Tropical Meteorology where some of this work was conducted. \\


\begin{thebibliography}{99}
\bibitem{foot0} If the initial conditions and/or forcing
project onto
a zonal/near-zonal modes then, much like the rapid dissipation of ageostrophic components in the 
3D rotating Boussinesq
equations \cite{Bartello},\cite{SS-gafd}, the (catalytic) $(f,s,f)$
interactions will lead to the dissipation
of non-zonal components by
sweeping them to
small scales, thereby giving the (mistaken) impression of spontaneous zonal flow generation.
\bibitem{foot-in} We are using the term "inertial-range" quite loosely in the present work.
\bibitem{footer} 
An intersting point raised by one of the reviewers concerns the time-scales involved in the emeregence 
of these coherent features. It is possible that the seemingly isotropic eddies for small $\alpha$ will at a 
later time deform to yield corresponding zonal flows. We do not see this happening for fairly long times. In fact, 
most of the evolution in the fields occurs while the Dirichelt quotient \cite{MW-book} decays rapidly in time, 
and these snapshots
are well after a levelling off of this quotient.
\end{thebibliography}
\end{document}